# Single Frame Deblurring with Laplacian Filters

Baran Ataman
Tampere Research Center
Huawei Technologies
Tampere, Finland
baran.ataman@huawei.com

Esin Guldogan
Tampere Research Center
Huawei Technologies
Tampere, Finland
esin.guldogan@huawei.com

*Abstract*—Blind single image deblurring has been a challenge over many decades due to the ill-posed nature of the problem. In this paper, we propose a single-frame blind deblurring solution with the aid of Laplacian filters. Utilized Residual Dense Network has proven its strengths in super-resolution task, thus we selected it as a baseline architecture. We evaluated the proposed solution with state-of-art DNN methods on a benchmark dataset. The proposed method shows significant improvement in image quality measured objectively and subjectively.

## I. INTRODUCTION

Images captured by digital cameras are usually suffered from blur. Two main types of blur are motion blur and out-of-focus blur, which result:

$$y = (x \otimes k) + n$$

where **x** represents the latent clear image, $\otimes$ is two-dimensional convolution, **y** represents the blurred image and **n** is the additive noise upon the captured image.
Blurred images can be classified as either globally blurred or locally blurred. Typical reasons for obtaining global blur are camera shake and out of focus, whereas local blur is caused by the moving scene/object when the camera is still and properly placed. Local blur can be identified with several different approaches ([7], [8], [9]) and we are focusing on single-frame local deblurring in this paper. We are proposing a network architecture in which both the locally blurred image and its Laplacian filtered version are fed so that the network pays attention to the edges of the original image.

## II. PREVIOUS WORKS

In this section, we briefly explain existing works on single-frame blind image deblurring methods in the literature:

### A. Multi-Frame Deblurring

Multi-frame deblurring methods exploit information from neighboring frames to resolve dynamic scene deblurring. In [10], a unified multi-image deconvolution algorithm is proposed with a special penalty function that couples the latent sharp image, blur kernels and noise variances. Cai *et al.* [12] identify the blur kernels and deblur the image by exploring the sparsity of the motion blur kernel and the clear image.
Recent CNN based methods achieved promising success for video deblurring. Su *et al.* [13] presented an algorithm in which 5 successive frames are fed into a CNN and the middle frame is deblurred after alignment of the frames with optical flow. A deep motion deblurring network proposed by [4] utilizes residual down-up and residual up-down blocks. It combines adaptive convolution results and residual images. Nah *et al.* [5] introduced a convolutional recurrent neural network architecture in order to exploit information among consecutive frames. They also employed intra-frame iterations which is basically an iterative hidden state update scheme within an inter-frame time step. As a result, their architecture is able to make use of both intra-frame and inter-frame schemes.

### B. Single-Frame Deblurring

Various solutions were proposed in order to uncover the latent clear image from a single blurry frame. Early research was focused on simultaneously estimating the blur kernel and the latent clear image by combining hand-crafted image priors [[14], [15], [16], [17]]. A recent approach proposed by Han and Kan [6] uses Laplacian operator to extract the gradients in the image because of its superiority of seizing details inside the images. The image and blurring kernel are again estimated in an alternating manner with an additional image mask to set a local constrain in order to eliminate ringing artifacts.
Recent developments in deep learning led to performance augmentations in single frame deblurring. Nah *et al.* [19] proposed an end-to-end deep learning approach which starts from a coarse scale of the blurry image and recovers the latent image at higher resolutions. Their model is trained with a multi-scale loss which leads to quick stable convergence. Kupyn *et al.* proposed a conditional generative adversarial network [21] architecture (DeblurGAN) [20] that utilizes gradient penalty [22] and perceptual loss [23], which empirically leads to sharper images with preserved texture details. DeblurGAN was superior to the previous ideas in motion deblurring with 5x more rapid than the fastest competitor. Residual Dense Network (RDN) [2], which is the inspiration of our architecture, has been the state-of-the-art in single-image-super-resolution (SISR) and it basically combines residual [26] and dense skip [27] connections. Although it was originally proposed for SISR task, the authors extended it and applied to restoration tasks [33]. A very recent approach called Self-Guided Network (SGN) [30] adopts a strategy to erase the noise in the image in a top-down approach. It uses large scale contextual information that is obtained from low spatial resolution to manage the feature extraction at finer scales. Their experiments illustrate that they serve equally favorable performance with the state of the art deblurring methods [31], [32] with less memory consumption and running time.

## III. PROPOSED METHOD

We utilized RDN architecture for deblurring tasks due to its local feature fusion and contiguous memory mechanism features. These features and their benefits are explained briefly in the next section. We added Laplacian filtered image as a second channel input for expressing the attention to the edges of the images. Local object blur degrades the edges of the objects within the images, on the other hand static objects and background stays sharp. Thus Laplacian filtered images can clearly show the degradation of the edges of moving objects.

Proposed method has the following 4 novelties:

- Modified version of RDN: RDN was originally proposed for single image super-resolution, our architecture is the suitable version of it for deblurring tasks.
- We feed RDN with the luminance (L) channel of the LAB color space instead of RGB channels in order to decrease the training and processing time. As indicated in [33], train and test time of RDN are significantly higher than most of the methods in the literature.
- Laplacian filter: In addition to the L channel of the deblurred images, Laplacian filtered versions are also given to the network for special attention to the areas closer to the edges.
- The loss function: A special loss function is proposed to preserve the edges of the blurred object. The proposed weighted loss function is explained in Section 3.D.

### A. Network Architecture

As shown in the Fig. 1, the architecture of our proposed network has 3 main parts: shallow feature extraction, residual dense blocks (RDB) and dense feature fusion. The first 2 convolution layers extract shallow features. $F_{-1}$ is used for another shallow feature extraction and global residual learning. $F_0$ is used as input to residual dense blocks. Assuming there are N number of RDBs, $F_N$ is obtained by

$$F_N = G_{RDB,N}(F_{D-1}) = G_{RDB,N}\left(G_{RDB,N-1}(...\left(G_{RDB,1}(F_0)\right)...)\right)$$

where $G_{RDB,d}$ denotes the operations of the $d^{th}$ RDB. $F_d$ is a local feature vector that utilizes each convolutional layers within the $d^{th}$ RDB. After extracting local features, the global feature vector $F_{GF}$ is obtained by global feature fusion [2]:

$$F_{GF} = G_{GFF}([F_1, ..., F_N])$$

where $G_{GFF}$ is the composition of $1 \times 1$ and $3 \times 3$ convolution and $[F_1, ..., F_N]$ is the concatenation of the local feature vectors produced by previous RDBs. $F_{GF}$ is formed by the adaptive fuse of local features coming from different levels.

The operation to obtain $F_{DF}$ is called global residual learning [2] and it is simply $F_{DF} = F_{GF} + F_{-1}$ where $F_{-1}$ is the shallow feature vector that is extracted at the beginning.

### B. Residual Dense Block

RDB architecture is depicted in Fig. 2. Densely connected layers, local feature fusion (LFF) and local residual learning lead to a contiguous memory (CM) mechanism [2].

Contiguous memory mechanism is the outcome of passing the state of the preceding block to each layer of the current RDB. $F_{M,c}$ denotes the resultant feature map of $c^{th}$ convolution layer of the $M^{th}$ RDB, and it is represented as

$$F_{M,c} = \sigma(W_{M,c}[F_{M-1}, F_{M,1}, ..., F_{M,c-1}])$$

where $\sigma$ is the activation function and $W_{M,c}$ is the learned weights of the $c^{th}$ convolutional layer. $F_{M,LF}$ is a local feature vector that is obtained by a $1 \times 1$ convolution to the concatenated feature maps that are produced by previous local feature extraction stages. $F_{M,LF}$ can be formulated as

$$F_{M,LF} = G_{LFF}^M([F_{M-1}, F_{M,1}, ..., F_{M,c}, ... F_{M,C}])$$

where $G_{LFF}^M$ represents the $1 \times 1$ convolution layer in the current RDB. This operation is called the local feature fusion [2].

Similar to global residual learning, local residual learning [2] is also implemented in order to enhance the information flow. The resultant vector of the $M^{th}$ RDB is computed as

$$F_M = F_{M,LF} + F_{M-1}$$

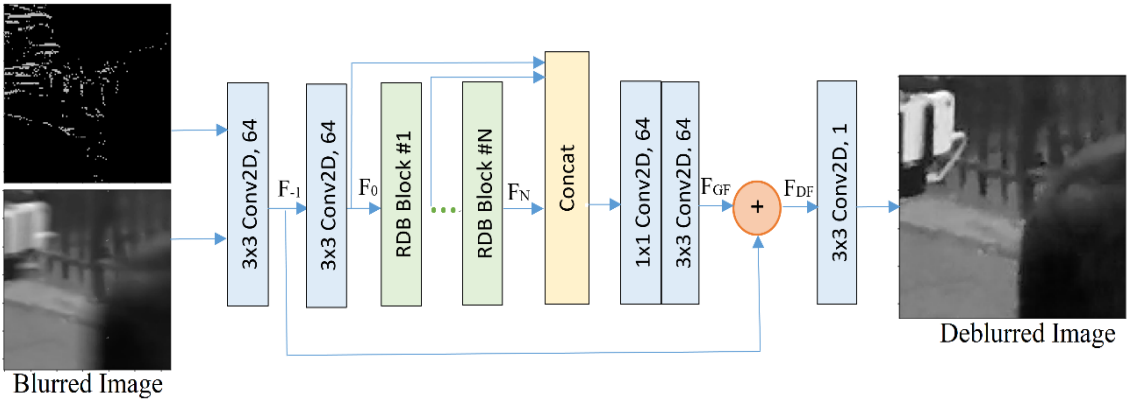

**Fig. 1**: The architecture of our proposed network

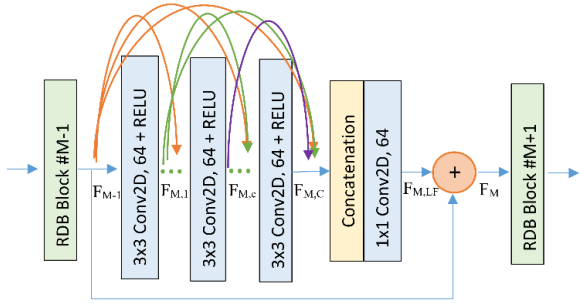

**Fig. 2:** Residual dense block (RDB) architecture

*C. Laplacian Filter*

The Laplacian filter calculates the second spatial derivative and is used to detect edges in the image. The Laplacian $L(x,y)$ can be calculated as follows:

$$L(x,y) = \frac{\partial^2 I}{\partial x^2} + \frac{\partial^2 I}{\partial y^2}$$

where $I$ is the intensity values of the image. $I$ is the set of discrete values, so we can easily calculate the Laplacian by convolving the input image with a small kernel. We selected the most common Laplacian kernel:

$$L = \begin{bmatrix} 0 & -1 & 0 \\ -1 & 4 & -1 \\ 0 & -1 & 0 \end{bmatrix}$$

We utilize the Laplacian filter image as an input in the training to enforce the network for increasing the attention to the regions where local blur occurs.

*D. Loss Function*

Common loss functions utilized in deblurring tasks are pixel-space loss e.g., the simplest L1 or L2 distance. However L1 and L2 losses tend to yield over smoothened pixel-space outputs and not discriminate successfully for the local blur artefacts. We proposed two-fold weighted loss function referred as weighted edge loss (WEL) which can be expressed as the following:

$$WEL = w_{l2} L2 + w_{el} EL$$

where EL is the edge preserving loss as the following:

$$EL = \sum_i^N (\frac{1}{N}) \|\nabla I_A - \nabla I_B\|$$

$\nabla I_A$ and $\nabla I_B$ denote the gradients of the input and ground truth images whereas $w_{l2}$ and $w_{el}$ represents coefficients.

## IV. EXPERIMENTAL RESULTS

*A. Training Details*

We trained our model on the GOPRO dataset [19], which includes 2103 training and 1111 test pairs. The pairs are realistic blurry and sharp images using a high-speed camera. We compare our results with those of the state-of-the-art methods in both qualitative and quantitative ways.

We selected 256 x 256 patch size and applied 4 times data augmentation by rotating the images around x and y axis as well as rotating 90 and 270 degrees. We had total of 520600 patches per each epoch compared to 16000 patches used by [33]. Batch size is selected as 4 and trained for 200 epochs. RDN is said to take 1 day for 200 epoch training with 16000 patches when RGB channels are used [33]. By using L channel with Laplacian filtered version, we decreased the training time of the model compared to the original RDN for deblurring [33] which uses RGB channels. Training took 20 days on a PC with 1 RTX 2080 GPU with Tensorflow libraries, whereas [33] theoretically takes 33 days with 520600 patches, so there is about 40% reduce in the training time in our RDN architecture. The optimizer was Adam [34] where Lr = 0.0001, β1 = 0.9, β2 = 0.999 and learning decay = 0.00005.

*B. Testing Details*

Since the authors of [33] did not share the training or the testing code of the deblurring architecture, we could not compare our method with the original RDN which is fed with 3 channels of RGB space. Instead we trained DeblurGAN [20] and SGN [30] with the same methodology (feeding with luminance channel of the LAB space) on the GoPro training set [18]. After the training, we compared these 2 recent state-of-the-art methods with our proposed network.

To evaluate the image quality we calculated PSNR, SSIM and MS-SSIM [18] metrics with ground-truth and deblurred images. Table 1 shows the average PSNR, SSIM and MS-SSIM values on GoPro test set which has 1111 images. As can be observed from the table, the proposed method overcomes the state-of-art methods on objective metrics. For a fair comparison we cropped the same 256 x 256 patches from test images and made the comparison on luminance channel. The outputs are deblurred luminance channels.

Fig. 3 shows example images from the GoPro test set. The proposed method removes the blur artefacts from the input images as well as providing the highest SSIM and MS-SSIM values. Proposed method gives satisfactorily outperforming results in case of visual comparison.

## V. CONCLUSION

We proposed residual dense network for single frame deblurring with Laplacian filters. Our primary additions were the augmented network architecture, reduced training time and a new loss function which is optimal for training the CNN. We compared our results with the state-of-the-art approaches of the recent years and the proposed method outperforms the compared methods in quantitative metrics. Further research is implementing and training the original RDN for RGB channels and compare it with our own method.

TABLE 1 OBJECTIVE METRIC COMPARISON ON GOPRO DATASET

| Method | PSNR | SSIM | MS-SSIM |
|---|---|---|---|
| DeblurGAN | **24.47** | 0.74 | 0.87 |
| SGN | 20.12 | 0.72 | 0.86 |
| Proposed | 23.39 | **0.85** | **0.93** |

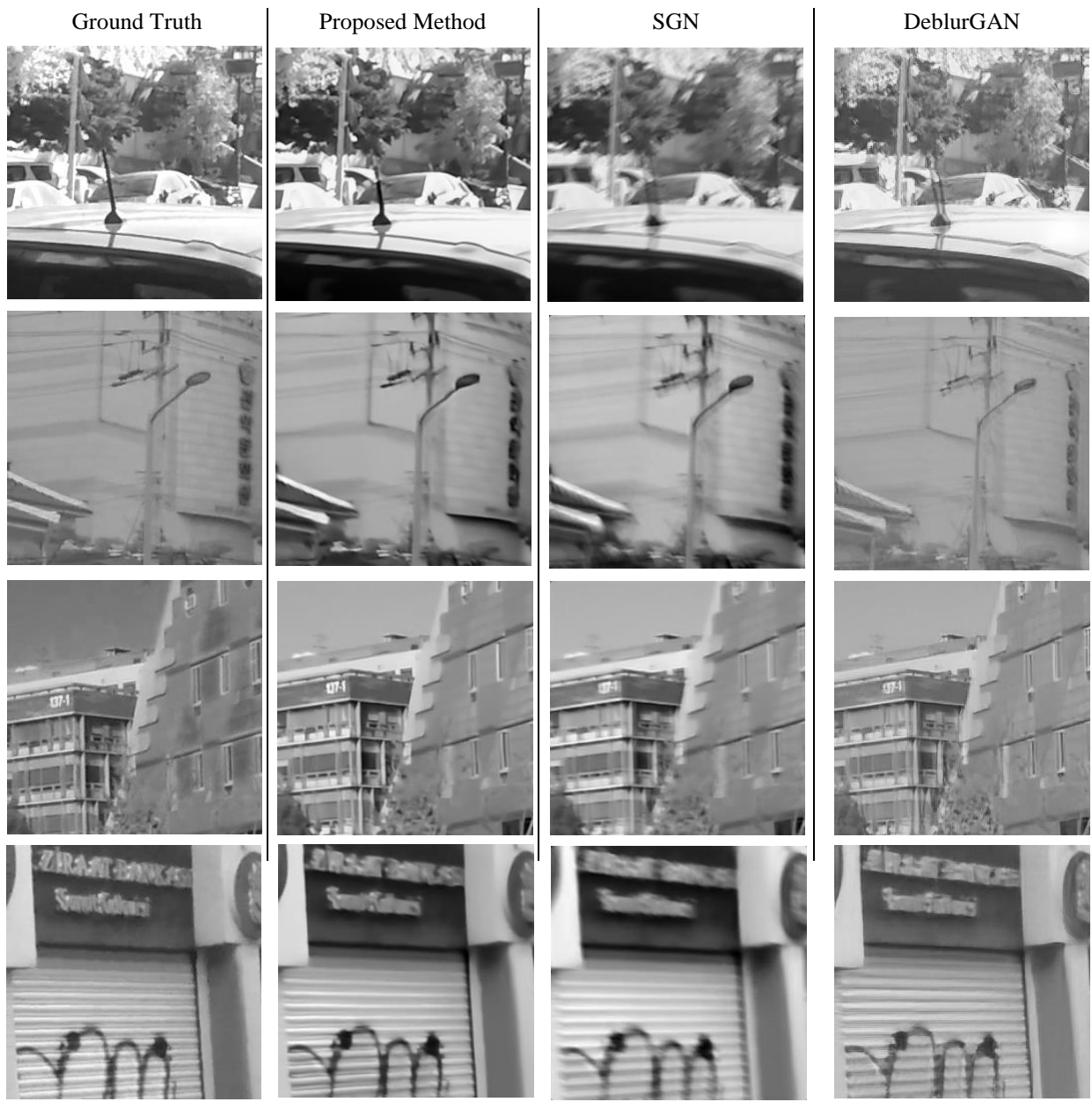

**Fig. 3:** Comparison of 3 methods on GoPro dataset [18] with the ground truth images